\pdfoutput=1
\documentclass{article}
\def\Caption{\caption}

\usepackage[T1]{fontenc}
\usepackage{palatino}
\usepackage{makeidx}
\makeindex
\usepackage{amsmath}
\usepackage{textcomp}% for upright mu (\textmu)
\usepackage{graphicx}
\usepackage{cite} % Compresses entries: [1,2,3] -> [1-3]

\begin{document}

%\setlength{\oddsidemargin}{0cm}
%\setlength{\evensidemargin}{0cm}
%\setlength{\topmargin}{0cm}
%\printindex\setcounter{page}{1}

\title{Microfabricated Chip Traps for Ions\footnote{This is a chapter from the forthcoming book ``Atom Chips''
edited by J. Reichel and V. Vuletic (to be published by WILEY-VCH).}}
\label{ions:chap}
\date{\small December 17, 2008}
\author{J. M. Amini, J. Britton, D. Leibfried, and D. J. Wineland\\
\small Time and Frequency Division\\
\small National Institute of Standards and Technology\\
\small Boulder, CO 80305}
\maketitle

\tableofcontents

%%%%%%%%%%%%%%%%%%%%%%%%%%%%%%%%%%%%%%%%%%%%%%%%%%%%%%%%%%%%%%%%%%%%%%%%%%%%%%%
%% INTRO
%%%%%%%%%%%%%%%%%%%%%%%%%%%%%%%%%%%%%%%%%%%%%%%%%%%%%%%%%%%%%%%%%%%%%%%%%%%%%%%
\section{Introduction}
\label{ions:sec:intro}

Most chapters of this monograph focus on trapping and manipulating neutral
 atoms with magnetic and optical fields. In this chapter, we discuss the trapping
 of atomic ions. This is of current high interest because individual ions can
 be the physical representations of qubits for quantum information processing
 \cite{ions:Cirac95.PRL.74.4091}. For recent reviews see 
 \cite{ions:Blatt08.Nature.453.1008,ions:Monroe08.PhysicsWorld.Aug}. The goals are similar to those of 
 neutral atom traps in that we wish to
 create microfabricated structures to trap, transport, and arrange ions in an array.
 Microfabrication holds the promise of forming large arrays of traps that would allow
 the scaling of current quantum information processing capabilities to the level 
 needed  to implement useful algorithms  
 \cite{ions:bible,ions:Kielpinksi02.Nature.417.709,ions:Kim05.QIC.5.515,ions:Steane07.QIC.7.171}.
 
 \index{ion trap}\index{Paul trap}\index{Penning trap}
 There are two primary types of ion traps used in low energy atomic physics: Penning traps and Paul traps. In a
 Penning trap, charged particles are trapped by a combination of static electric and
 magnetic fields \cite{ions:Penning36.Scripta.3.873,ions:Dehmelt90.RMP.62.525}.  In a Paul trap, 
 a spatially varying 
 sinusoidally oscillating electric field, typically in the radio-frequency (rf)
 domain, confines atomic or molecular ions in space \cite{ions:Paul90.RMP.62.531}. In this review
 only the Paul type will be considered.
 
 Neutral atom traps operate by a coupling between external trapping fields and
 atoms' electric or magnetic moments. Trap depths of
 a few kelvins are common.  In ion traps, an ion is trapped by a coupling between the applied
 electric trapping fields and the atom's net (or overall) charge.  
 Typical ion trap depths are 1~eV. This coupling
 does not depend on the ion's internal electronic state, leaving it largely unperturbed.

 We begin this chapter with an introduction to the dynamics 
 of ions confined in Paul traps
based on the pseudopotential approximation.  Subsequent topics include numeric
and analytic models for various Paul trap geometries, a list of considerations
for practical trap design and finally an overview of microfabricated trapping
structures. A discussion of future directions concludes this chapter.

% %%%%%%%%%%%%%%%%%%%%%%%%%%%%%%%%%%%%%%%%%%%%%%%%%%%%%%%%%%%%%%%%%%%%%%%%%%%%%%
% %
% %%%%%%%%%%%%%%%%%%%%%%%%%%%%%%%%%%%%%%%%%%%%%%%%%%%%%%%%%%%%%%%%%%%%%%%%%%%%%%
\section{Radio-frequency (rf) ion traps}
\label{ions:sec:ModOfR}

\index{ions!trapping}\index{rf trap} In this section we discuss the equations of motion of a charged
particle in a spatially inhomogeneous radio-frequency (rf) field based on the pseudopotential approximation model.
We then present examples of suitable electrode geometries.

% %%%%%%%%%%%%%%%%%%%%%%%%%%%%%%%%%%%%%%%%%%%%%%%%%%%%%%%%%%%%%%%%%%%%%%%%%%%%%%
% %
% %%%%%%%%%%%%%%%%%%%%%%%%%%%%%%%%%%%%%%%%%%%%%%%%%%%%%%%%%%%%%%%%%%%%%%%%%%%%%%
\subsection{Motion of ions in a spatially inhomogeneous rf field}
\label{ions:subsec:EquMot}\index{Paul trap!ion motion}

\begin{figure}[htb]
\centering
  \includegraphics{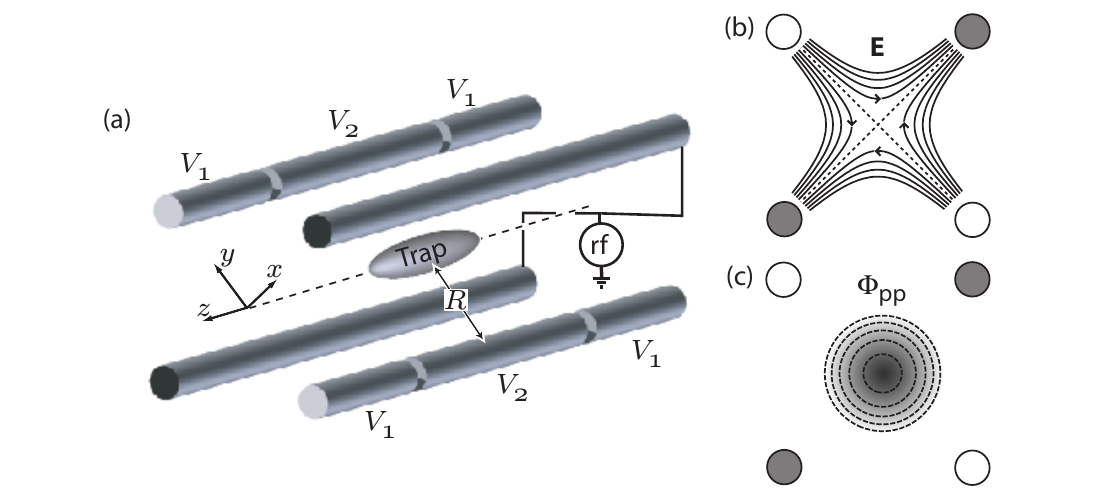}
  \Caption{ (a) Schematic drawing of the electrodes for a linear Paul trap. A common
  rf~potential $V_0\cos(\Omega_{\rm rf}t)$ is applied to the two
  continuous electrodes, as indicated. The other electrodes are held at rf ground
  through capacitors (not shown) connected to ground. In (b), we show
  the radial ($x$-$y$) instantaneous electric fields from the applied rf potential. Contours
  of the pseudopotential due to this rf-field are shown in (c). 
  A static trapping potential is created  along
  the z-axis by applying a positive potential $V_1>V_2$ (for positive ions) 
  to the outer segments relative to the center segments.}
  \label{ions:fig:LinEle}
\end{figure}

Most schemes for quantum information processing with trapped ions are
based on a linear rf trap shown schematically in
fig.~\ref{ions:fig:LinEle}a. This trap is essentially a linear quadrupole mass
filter \cite{ions:Paul90.RMP.62.531} with its ends plugged by static potentials \cite{ions:Wineland04.QEIP.261}. 
The radial confinement (the $x$-$y$ plane in fig.~\ref{ions:fig:LinEle}a) is
provided by an rf potential applied to 
two of the electrodes  with the other electrodes held at rf ground. 
In this linear geometry, the rf potential cannot generate full 3D confinement, 
so static potentials $V_1$ and $V_2$ applied to control electrodes provide
axial ($z$ axis) confinement.  We will assume the axial trapping fields are relatively
weak so that the accompanying static radial fields do not significantly perturb the radial trapping.

Applying a potential of $V_0\cos(\Omega_{\rm rf} t)$ to the rf electrodes while
grounding the other electrodes ($V_1=V_2=0$), the rf potential near the geometric center
of the four rods takes the form
\begin{equation}\label{ions:eq:fullPotential}
  \Phi\approx\frac{1}{2}V_0\cos(\Omega_{\rm rf}
  t)(1+\frac{x^2-y^2}{R^2}),
\end{equation}
where $R$ is a distance scale that is approximately the distance from the trap
axis to the nearest surface of the electrodes
\cite{ions:Paul90.RMP.62.531,ions:bible,ions:Wineland04.QEIP.261}. The resulting electric field is
shown in fig.~\ref{ions:fig:LinEle}b. There is a field null at the trap center;
 the field magnitude increases linearly with distance from the center.

We can think of the rf electric field as analogous
to the electric field from the trapping laser in an optical dipole
trap \cite{ions:bible,ions:grimm2000a}.  For a neutral atom, the laser's electric
field induces a dipole moment. If the electric field is inhomogeneous, 
the force on the dipole, averaged over one cycle of the radiation, can give a trapping force.
For detunings red of the atom's resonant frequency $\omega_0$, the resulting potential
is a minimum at high fields, while for detunings blue of $\omega_0$ it is a minimum at low
fields. An ion, however, is a free particle in the absence of a trapping field
and its eigenfrequency is zero. The rf trapping
potential is therefore analogous to a blue detuned light field and the ion seeks the position
of lowest intensity. In the case of eq.~(\ref{ions:eq:fullPotential}), that corresponds to
$x=y=0$.

The motion for an ion placed in this field is commonly treated in
one of two ways: as an exact solution of the \index{Mathieu equation} Mathieu differential equation or as an approximate solution
of a static effective potential called the `pseudopotential'. The 
Mathieu solutions provide insights
on trap stability and high frequency motion; the pseudopotential approximation
is more straightforward and is convenient for the analysis of trap designs. 

\index{pseudopotential} We define the pseudopotential that governs the \index{secular motion} secular motion as follows \cite{ions:Dehmelt67.AAMP.3.53}. 
The motion of an ion in the rf field is a combination of fast \index{micromotion} `micromotion' at
the rf frequency on top of a slower `secular' motion.
For a particle of charge $q$ and mass $m$ in a
uniform electric field $E=E_0\cos(\Omega_{\rm rf} t)$, the ion motion (neglecting a drift term) takes the form
\begin{equation} \label{ions:eq:ionmotion}
  x(t)=-x_{\mu m} \cos(\Omega_{\rm rf} t),
\end{equation}
where $x_{\mu m}=q E_0/(m \Omega_{\rm rf}^2)$ is the amplitude of what we will call micromotion.
If the rf field amplitude has a spatial dependence $E_0(x)$ along the $x$ direction, there is a nonzero
net force on the ion when we average over an rf cycle:
\begin{equation}
  F_{\rm net} = \left<qE(x)\right> \approx -\frac{1}{2}q\left.\frac{dE_0(x)}{dx}\right|_{x\rightarrow x_s}x_{\mu m}
  = -\frac{q^2}{4m\Omega_{\rm rf}^2}\left.\frac{dE_0^2(x)}{dx}\right|_{x\rightarrow x_s}
  = -\frac{d}{dx} (q\Phi_{\rm pp}),
\end{equation}
where $x$ is evaluated at what we designate as the secular position $x_s$, and the pseudopotential 
$\Phi_{\rm pp}$ is defined by
\begin{equation}\label{ions:pseudopotential}
  \Phi_{\rm pp}(x_s)\equiv\frac{1}{4}\frac{q E_0^2(x_s)}{m \Omega_{\rm rf}^2}.
\end{equation}
We have made the approximation
that the solution in eq.~(\ref{ions:eq:ionmotion}) holds over an rf cycle and have
dropped terms of higher order in the Taylor expansion of $E_0(x)$ around $x_s$.  For regions
near the center of the trapping potential, these approximations hold.
In three dimensions, we make the substitution $E_0^2 \rightarrow |E|^2=E_{0,x}^2+E_{0,y}^2+E_{0,z}^2$. 
Note that the pseudopotential depends on the magnitude of the
electric field, not its direction. 

%The pseudopotential has a second interpretation in terms of conservation of energy.
%From eq.~(\ref{ions:eq:ionmotion}), the kinetic energy averaged over a single rf cycle is 
%\begin{equation}
%  \mathcal{E}_{\rm total} = q\Phi_{\rm pp} + \frac{1}{2}m v_s^2
%\end{equation}
%where the first term is the average kinetic energy of the micromotion and the second term
%is the kinetic energy of the secular motion. The pseudopotential therefore
%represents the average kinetic energy of the micromotion.  In the limit where the 
%approximations we made are valid, this total energy remains constant: any increase (decrease) in the secular
%kinetic energy is accompanied by a corresponding decrease (increase) in the average
%micromotion kinetic energy. 

For the quadrupole field given in eq.~(\ref{ions:eq:fullPotential}), the pseudopotential is that
of a 2D harmonic potential (see fig.~\ref{ions:fig:LinEle}c):
\begin{equation}\label{ions:eq:harmpot}
q\Phi_{\rm pp} = \frac{1}{2}m\omega_r^2(x^2+y^2),
\end{equation}
where \index{radial frequency} $\omega_r \simeq qV_0/(\sqrt{2}m\Omega R^2)$ is the resonant frequency. 
As an example, for $^{24}Mg^+$ in a Paul trap with $V_0=50$~V, $\Omega_{\rm rf}/2\pi=100$~MHz and
$R=50$~{\textmu m}, which are typical parameters for a microfabricated trap, the radial oscillation 
frequency is $\omega_r/2\pi=14$~MHz. 

The rf pseudopotential provides confinement of
the ion in the radial ($x$-$y$) plane. Axial trapping is obtained by the addition of the static
control potentials $V_1$ and $V_2$, as shown in fig.~\ref{ions:fig:LinEle}a.  

For $\omega_z \ll \omega_r$, multiple ions trapped in the same potential well 
will form a linear `crystal' along the trap axis due to  a balance between the axial trapping potential and the ions'
mutual Coulomb
repulsion. \index{ions!spacing} The inter-ion spacing is determined by the axial frequency
($\omega_z$). The characteristic length scale of ion-ion spacing is
\begin{equation}\label{ions:eq:ionspacing}
  s=\left(\frac{q^2}{4 \pi \epsilon_0 m \omega_z^2}\right)^{1/3}.
\end{equation}
For a three-ion crystal the adjacent separation of the ions is
$s_{3}=(5/4)^{1/3}s$
\cite{ions:bible}. For example, $s_3=5.3$~{\textmu m} for $\,^{24}\text{Mg}^{+}$ and
$\omega_{z}/2\pi=1.0$~MHz. For multiple ions in a linear Paul trap,
$\omega_z$ is the frequency of the lowest vibrational mode  (the center of mass mode) along the trap axis.

A single ion's radial  motion  in the potential given by eq.~(\ref{ions:eq:harmpot}) can be decomposed 
into uncoupled harmonic motion in the $x$ and $y$ directions, both with the same trap frequency $\omega_r$.  
Because the potential is cylindrically symmetric
about $z$, we could choose the decomposition about any two orthogonal directions, called the principle
axes.  We will see in section~\ref{ions:sec:doppler} when discussing Doppler cooling 
that we need to break this cylindrical symmetry by the application of static electric fields. In that
case, the choice of the principle axes becomes fixed with corresponding radial trapping frequencies $\omega_1$
and $\omega_2$, one for each principle axis.     

% %%%%%%%%%%%%%%%%%%%%%%%%%%%%%%%%%%%%%%%%%%%%%%%%%%%%%%%%%%%%%%%%%%%%%%%%%%%%%%
% %
% %%%%%%%%%%%%%%%%%%%%%%%%%%%%%%%%%%%%%%%%%%%%%%%%%%%%%%%%%%%%%%%%%%%%%%%%%%%%%%
\subsection{Electrode geometries for linear quadrupole traps}
\label{ions:sec:EleGeo} \index{Paul trap!geometries}

\begin{figure}[htb]
\centering
  \includegraphics{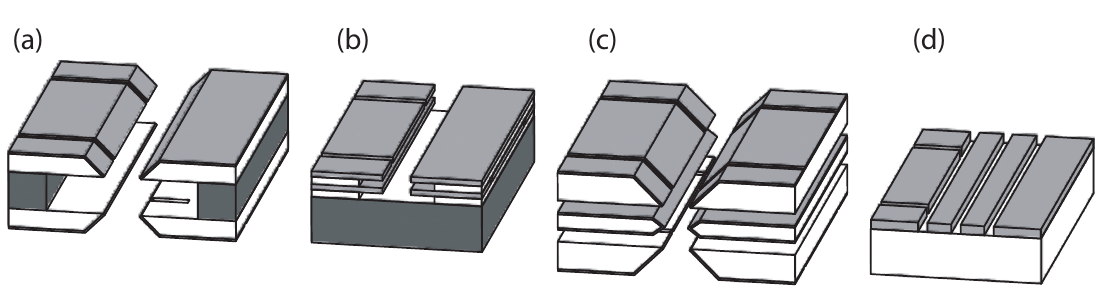}
  \Caption{Examples of microfabricated trap structures: (a) two wafers mechanically
  clamped over a spacer 
  \cite{ions:Rowe02.QIC.2.257, ions:Wineland05.ICOLS.393,  ions:Huber08.NJP.10.013004, ions:Schulz08.NJP.10.045007, ions:blakestad2008a}, 
  (b) two layers of electrodes fabricated onto a single wafer
  \cite{ions:Stick06.NaturePhys.171.36}, 
  (c) three wafers clamped with spacers (not shown)
  \cite{ions:Hensinger06.APL.88.034101}, 
 	and (d) surface electrode construction
 	\cite{ions:Chiaverini05.QIC.5.419, ions:Pearson06.PRA.73.032307, ions:Seidelin06.PRL.96.253003, ions:Britton06.ArXiv.0605170, ions:Brown07.PRA.75.015401, ions:Labaziewicz08.PRL.100.013001, ions:Britton08.Thesis}.}
  \label{ions:fig:examples}
\end{figure}

Designs for miniaturized ion traps conserve the basic features of the Paul 
trap shown in fig.~\ref{ions:fig:LinEle}. Figure~\ref{ions:fig:examples} shows  a few geometries that
have been experimentally realized. All these geometries  generate 
a radial quadratic potential near the trap axis, though the extent of deviations from 
the ideal quadrupole potential away from the axis will depend on the design.  

\index{surface electrode (SE) trap}In one particular geometry, the electrodes all lie in
a single plane, as shown in fig.~\ref{ions:fig:examples}d with the ion suspended above the plane 
\cite{ions:Chiaverini05.QIC.5.419,ions:Pearson06.PRA.73.032307,ions:Seidelin06.PRL.96.253003,ions:Britton06.ArXiv.0605170,ions:Brown07.PRA.75.015401,ions:Labaziewicz08.PRL.100.013001,ions:Britton08.Thesis}.
Trapping in such surface electrode (SE) traps is possible over a wide range of geometries, albeit
with $1/6$ to $1/3$ the motional frequencies and $1/30$ to $1/200$ the trap depth
 of more conventional quadrupolar geometries at comparable rf potentials and 
ion-electrode distances \cite{ions:Chiaverini05.QIC.5.419}. 

Advantages of the SE trap geometry over the other geometries shown in fig.~\ref{ions:fig:examples}
include easier fabrication and the possibility
of integrating control electronics on the same trap wafer \cite{ions:Kim05.QIC.5.515}.
A SE trap at cryogenic temperature was demonstrated at MIT in 2008 \cite{ions:Labaziewicz08.PRL.100.013001}.  

Research on SE trap designs
is ongoing and holds promise to yield complex geometries that would be difficult
to realize in non-surface electrode designs.

%%%%%%%%%%%%%%%%%%%%%%%%%%%%%%%%%%%%%%%%%%%%%%%%%%%%%%%%%%%%%%%%%%%%%%%%%%%%%%%
%%
%%%%%%%%%%%%%%%%%%%%%%%%%%%%%%%%%%%%%%%%%%%%%%%%%%%%%%%%%%%%%%%%%%%%%%%%%%%%%%%
\section{Design considerations for Paul traps}
\label{ions:sec:considerations}
In this section, we will discuss the requirements that need to be addressed 
when designing a practical ion trap. 

%%%%%%%%%%%%%%%%%%%%%%%%%%%%%%%%%%%%%%%%%%%%%%%%%%%%%%%%%%%%%%%%%%%%%%%%%%%%%%%
%%
%%%%%%%%%%%%%%%%%%%%%%%%%%%%%%%%%%%%%%%%%%%%%%%%%%%%%%%%%%%%%%%%%%%%%%%%%%%%%%%
\subsection{Doppler cooling}
\label{ions:sec:doppler}\index{Doppler cooling}\index{ions!Doppler cooling}

For Doppler laser cooling of an ion in a trap, only a single laser beam is needed; trap
strengths far exceed the laser beam radiation pressure. 
The cooling is offset by heating from photon recoil. Therefore, to cool in all
directions, the Doppler cooling beam  k-vector must have a component along 
all three principal axes of the trap \cite{ions:Itano82.PRA.25.35}.  This also implies that the trap frequencies are 
not degenerate, otherwise one principal axis could be chosen normal to the laser beam's k-vector.

\begin{figure}[htb]
\centering
\includegraphics{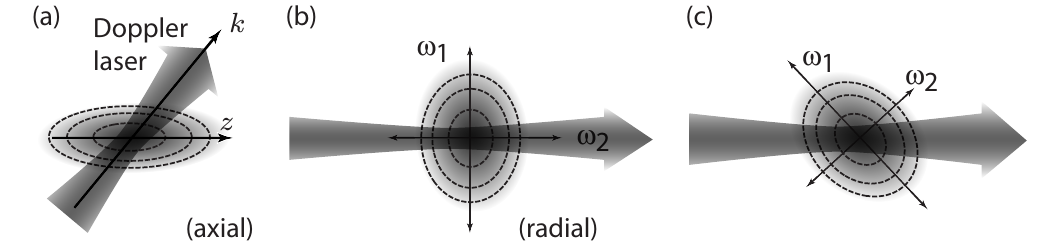}
\Caption{Doppler cooling with a single laser beam.  The dashed
lines are equipotential curves for the pseudopotential. The overlap with the
axial direction ($z$) is fairly straightforward, as in (a), but care has to be taken that the
orientation of the two radial modes $\omega_1$ and $\omega_2$ does not place one of the mode axes
perpendicular to the laser beam, as shown in (b). For efficient cooling, the axes must be at an angle 
with respect to the laser beam k-vector (c).} % Note the capital "C" in "\Caption"
\label{ions:fig:cooling}
\end{figure}

Meeting the first condition is usually straightforward for non-SE type traps, where access for the laser beam
is fairly open (see fig.~\ref{ions:fig:cooling}). For SE traps, where laser beams are 
typically constrained to run parallel to the chip surface,  care has to be taken in 
designing the trap so that neither radial principle axis is perpendicular to the trap surface.
Alternately, for SE traps, we could bring the Doppler laser beam at an angle to the surface but
the beam would have to strike the surface.  This can cause problems with scattered light affecting
\hyphenation{fluo-rescence} detection of the ion and with charging of exposed dielectrics (see 
section~\ref{ions:sec:exposedDielectric}).

\begin{figure}[htb]
\centering
\includegraphics{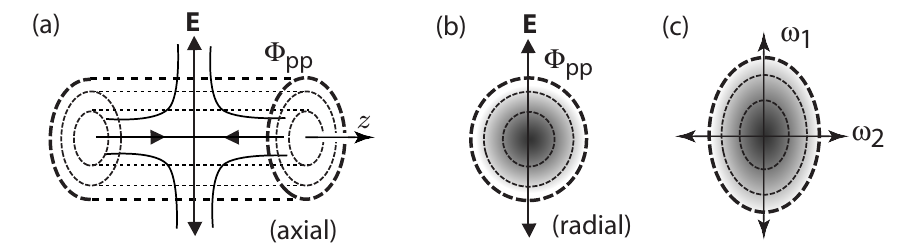}
\Caption{The degeneracy in the radial trap modes can be lifted by the radial
component of the static axial confinement field. In (a), the quadrupole field is shown
overlaid on the cylindrically symmetric pseudopotential.  The radial
component of the electric field (b) deforms the net potential seen by the ion
(c), breaking the cylindrical symmetry.} % Note the capital "C" in "\Caption"
\label{ions:fig:trap_rotation}
\end{figure}

If any two trap frequencies are \index{degenerate frequencies} degenerate, then the trap axes  in the plane containing those modes 
are not well defined and
the motion in a direction perpendicular to the Doppler laser beam k-vector will not be cooled
and will be heated due to photon recoil.  
The axial trap frequency can be set independently of the radial frequencies and
can be chosen to prevent a degeneracy with either of the radial modes. 
However, the two radial modes could still be degenerate. 
There are several ways to break this degeneracy, but usually the axial trapping potential
is sufficient.  When we apply an axial
trapping potential, Laplace's equation forces us to have a radial component to the
electric field. In general, this radial field is
not cylindrically symmetric about $z$ and will distort the net trapping
potential, as shown in fig.~\ref{ions:fig:trap_rotation}, thereby lifting the
degeneracy of the radial frequencies. If this is not
sufficient,  offsetting {\it all} the control electrodes by a common potential  with respect to the rf electrodes
will result in a static field that has the same spatial dependence 
(that is the same function of $x$ and $y$) as the
field generated by the rf electrodes. This field, shown in fig.~\ref{ions:fig:LinEle}b, 
can be used to split the radial frequencies. We will refer to the axes' orientation
resulting from the offset of all control electrodes as the \index{intrinsic axes} `intrinsic' trap axes since
it does not depend on the segmentation of the control electrodes, but only on the overall geometry
of the rf and control electrodes.
The static axial potential might or might not define trap axes aligned with the intrinsic axes,
 but, overall, the control electrodes and axial potential can be configured to prevent either
 radial modes from being normal to the surface in an SE trap.
Furthermore, in some cases additional control electrodes are designed into the trap 
to lift the degeneracy independent of both the axial potential and the intrinsic axes.

%%%%%%%%%%%%%%%%%%%%%%%%%%%%%%%%%%%%%%%%%%%%%%%%%%%%%%%%%%%%%%%%%%%%%%%%%%%%%%%
%%
%%%%%%%%%%%%%%%%%%%%%%%%%%%%%%%%%%%%%%%%%%%%%%%%%%%%%%%%%%%%%%%%%%%%%%%%%%%%%%%
\subsection{Micromotion}
\label{ions:sec:micromotion}\index{micromotion}

If the pseudopotential at the equilibrium position of a trapped ion is nonzero,
then the ion motion will include a persistent micromotion component at frequency $\Omega_{\rm rf}$.  
There are two mechanisms that can
generate a nonzero equilibrium pseudopotential. 
As the trapping structures become more complicated and the
symmetry of the simple Paul trap in fig.~\ref{ions:fig:LinEle} is broken, 
there can be a component of the rf field in the axial direction at 
the pseudopotential minimum; that is, the pseudopotential minimum 
need not be a pseudopotential zero. Since
this effect is caused by the geometry of the trap, we refer to the resulting micromotion 
as `intrinsic' micromotion \cite{ions:Berkeland98.JAP.83.5025}.
Secondly, if there is a static electric field at the pseudopotential zero, 
the equilibrium position
of an ion will be shifted away from the pseudopotential minimum.  Because shim
potentials can be applied to the control electrodes to null these fields \cite{ions:Berkeland98.JAP.83.5025},
the micromotion due to this mechanism is called `excess' micromotion.

\begin{figure}[htb]
\centering
\includegraphics{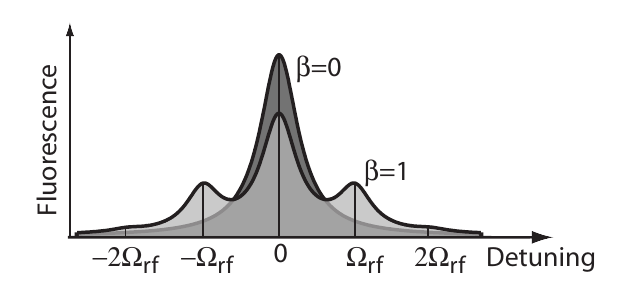}
\Caption{In the rest frame of the ion, micromotion induces sidebands of a probing
laser.  Here, the monochromatic laser spectrum has been convoluted with the atomic
linewidth.}
\label{ions:fig:sidebands}
\end{figure}

Both intrinsic and excess micromotion can cause problems 
with the laser-ion interactions, such as Doppler cooling,
ion fluorescence, and Raman transitions \cite{ions:bible,ions:Berkeland98.JAP.83.5025}.
An ion with micromotion experiences a frequency-modulated laser field due to the Doppler shift. In the rest frame of the ion, 
this modulation introduces \index{sidebands} sidebands to the laser frequency (as seen by the ion) at
integer multiples of $\Omega_{\rm rf}$ and reduces the laser beam's intensity at the carrier frequency,
as shown in fig.~\ref{ions:fig:sidebands}.  The strength of these sidebands is 
parametrized by the modulation index $\beta$, given by
\begin{equation}
\beta = \frac{2\pi x_{\mu m}}{\lambda} \cos \theta
\end{equation}
where $x_{\mu m}$ is the micromotion amplitude, $\lambda$ is the laser wavelength, and $\theta$ is the angle the laser 
 beam k-vector makes with the micromotion.  For laser beams tuned near resonance, ion fluorescence becomes 
weaker and can disappear entirely.
As another example, when $\beta=1.43$, the carrier and first micromotion sideband have equal strength.  
For $\beta<1$, the fractional loss of on-resonance fluorescence is approximately $\beta^2/2$. 
As a rule of thumb, we aim for $\beta<0.25$, which corresponds to a drop of less than five percent in
on-resonant fluorescence.

For a given static electric field $E_{\rm dc}$ in the radial plane,  
an ion's radial displacement $x_d$ from the trap center
and the resulting excess micromotion amplitude $x_{\mu m}$ are
\begin{equation}
  x_d=\frac{q E_{\rm dc}}{m \omega_r^2},\\
  x_{\mu m} \simeq \sqrt{2}\frac{w_r}{\Omega_{\rm rf}}x_d,
\end{equation}
where $\omega_r$ is the radial trapping frequency. 

Assume $^{24}Mg^+$, $\Omega_{\rm rf}/2\pi=100$~MHz and $\omega_r/2\pi=10$~MHz.
A typical SE trap with $R\sim 50$~{\textmu m} and an excess potential of 1~V on a control
electrode will produce a radial electric field at the ion of $\sim500$~V/m.  The
resulting displacement is $x_d=500$~nm and the corresponding micromotion amplitude
is $x_{\mu m}=70$~nm.  This results in a laser modulation index of $\beta=1.14$.

Stray electric fields can be nulled if the control electrode geometry
permits application of independent compensation fields along each radial
principle axis. For the Paul trap in fig.~\ref{ions:fig:LinEle}, a common potential applied to
the control electrodes can only generate a field at the trap center that is
along the diagonal connecting the electrodes.
We can compensate for other directions by applying, for example,
 a static potential offset to one of the rf electrodes or by adding extra compensation electrodes.

There are several experimental approaches to detecting and minimizing excess micromotion
~\cite{ions:Berkeland98.JAP.83.5025}. One technique uses the dependence of 
the fluorescence from a cooling laser beam on the micromotion modulation index. The micromotion
can be minimized by maximizing the fluorescence when the laser is near resonance and 
minimizing the fluorescence when tuned to the rf sidebands. 

Intrinsic micromotion can also 
be caused by an rf phase difference
$\phi_{\rm rf}$ between the two rf electrodes. A phase difference can arise due to a path
length difference or a differential capacitive coupling to ground for the leads
supplying the electrodes with rf potential~\cite{ions:Berkeland98.JAP.83.5025,ions:Britton08.Thesis}. We aim for
$\beta<0.25$ (see section~\ref{ions:sec:micromotion}) for typical parameters,
which requires $\phi_{\rm rf}<0.5^\circ$.

%%%%%%%%%%%%%%%%%%%%%%%%%%%%%%%%%%%%%%%%%%%%%%%%%%%%%%%%%%%%%%%%%%%%%%%%%%%%%%%
%%
%%%%%%%%%%%%%%%%%%%%%%%%%%%%%%%%%%%%%%%%%%%%%%%%%%%%%%%%%%%%%%%%%%%%%%%%%%%%%%%

\subsection{Exposed dielectrics}
\label{ions:sec:exposedDielectric}\index{dielectric charging}

Exposed dielectric surfaces near the trapping region can pose a problem due
to charging of these surfaces  and resulting stray electric fields.  Charging can be caused by photo-emission by 
the probe laser or from electron sources such as those used 
for loading ions into the traps. Depending on the resistivity of
the dielectric, these charges can remain on the surfaces for minutes or longer, requiring
time-dependent micromotion nulling or waiting a sufficient time for the charge to dissipate.

Surface electrode traps can be particularly prone to this problem. The metallic trapping electrodes are 
often supported by an insulating substrate and the spaces between the electrodes
expose the substrate.  The effect of charging these regions can be mitigated by 
increasing the ratio of electrode conductor thickness to the inter-electrode spacing.  

\begin{figure}
\centering
  \includegraphics{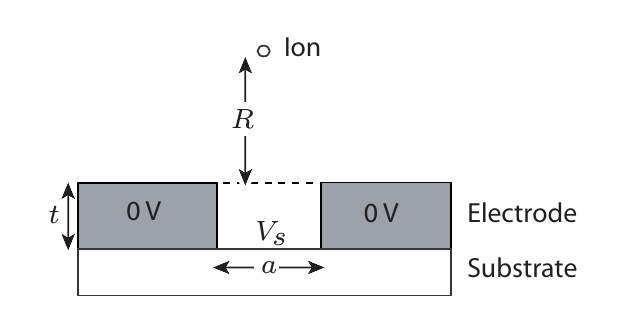}
  \Caption{Model used for estimating the effect of stray charging. We assume
  that $R\gg a$ and $t\geq a/\pi$.}
  \label{ions:fig:straycharge}
\end{figure}

Figure~\ref{ions:fig:straycharge} illustrates a model
for estimating how thick electrodes can suppress the field from a strip
of exposed substrate charged to a potential $V_s$.  
The sidewalls are assumed conducting and grounded.  Along the
midpoint of the trench the potential drops exponentially with height 
\cite{ions:jackson1999a}. Using this solution to relate $V_s$ to
the potential at the top of the trench, and employing the techniques
described in section~\ref{ions:sec:analyticsolutions} to
relate the surface potential to a field at the ions,
we obtain an approximate expression for the field seen by the ion:
\begin{equation}
|E| \simeq \frac{aV}{\pi R^2}\times
				\left\{ 
							\begin{array}{ll}
								1, & t=0 \\
								\frac{4}{\pi} e^{-\pi t/a}, & t\geq \frac{a}{\pi}, \\
							\end{array}
				\right.
\end{equation}
where $a$ is the width of the exposed strip of substrate, $t$ is the
electrode thickness, $R$ is the distance from the trap surface to the ion, 
and we have assumed $R\gg a$.  Thus, the effect
of the stray charges drops off rapidly with the ratio of
electrode thickness to gap spacing. 
  
%%%%%%%%%%%%%%%%%%%%%%%%%%%%%%%%%%%%%%%%%%%%%%%%%%%%%%%%%%%%%%%%%%%%%%%%%%%%%%%
%%
%%%%%%%%%%%%%%%%%%%%%%%%%%%%%%%%%%%%%%%%%%%%%%%%%%%%%%%%%%%%%%%%%%%%%%%%%%%%%%%
\subsection{Loading ions}
\label{ions:sec:loading} \index{ions!loading}

Ions are loaded into traps by ionizing neutral atoms as they pass through the
trapping region.  
The neutral atoms are usually supplied by a heated oven but can
also come from background vapor in the vacuum or laser ablation of a sample.

It is necessary that the neutral atom flux reach the trapping region but not deposit 
on insulating spacers, which might cause shorting between adjacent trap electrodes.  In practice, 
this is accomplished by careful shielding and, in some SE traps, undercutting of electrodes to form
a shadow mask (see fig.~\ref{ions:fig:signefab}).  
Alternately, for SE traps, a hole machined through the substrate can be used to
direct neutral flux from an oven on the back side of the wafer
to a small region of the trap, preventing coating of the surface.
This is called backside loading and has been demonstrated in several traps (see
section~\ref{ions:sec:traps}).

% %%%%%%%%%%%%%%%%%%%%%%%%%%%%%%%%%%%%%%%%%%%%%%%%%%%%%%%%%%%%%%%%%%%%%%%%%%%%%%
% %
% %%%%%%%%%%%%%%%%%%%%%%%%%%%%%%%%%%%%%%%%%%%%%%%%%%%%%%%%%%%%%%%%%%%%%%%%%%%%%%
\subsection{Electrical connections}
\label{ions:sec:elecInterconnect}

The control potentials and rf trapping potentials are delivered to
the trap electrodes by wiring that includes conducting traces on the trap substrate.  Care is
needed to avoid several pitfalls.

\index{resonator} The high-voltage rf potential is typically produced with resonant rf transformers \cite{ions:Macalpine59.PoIRE.59,ions:Cohen65.MWJ.8.69,ions:Jefferts95.PRA.51.3112}.
Rf losses in  a microtrap's electrodes or insulating
substrate can degrade the resonator (loaded) quality factor ($Q_L$) and can cause ohmic
heating of the microtrap itself.  This can be be mitigated by use of low-loss
insulators (for example, quartz or alumina) and decreasing the capacitive coupling of the
rf electrodes to ground through the insulators.  Typical rf parameters 
are $\Omega_{\rm rf}/{2\pi}=10$ to 100~MHz, $V_{\rm rf}\simeq100~V$ and $Q_L\simeq200$.

\begin{figure}
  \centering
  \includegraphics{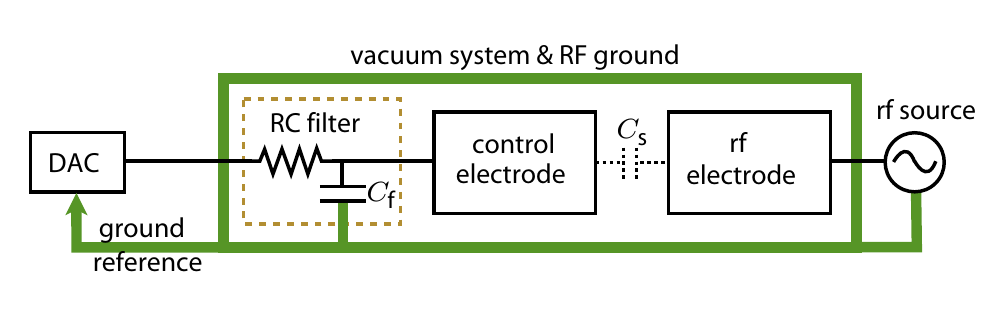}
  \Caption{Figure showing typical filtering and grounding of a trap control
  electrode. Inside the vacuum system are low pass RC filters which reduce noise
  from the control potential source and provide low impedance
  shorts to ground for the rf coupled to the control electrodes by stray 
  capacitances $C_{\rm s} \ll C_{\rm f}$. The RC filters typically lie inside the vacuum
  system, within 2~cm of the trap electrodes. The control potential
  is referenced to the trap rf ground and is supplied over a properly  shielded wire.}
\label{ions:fig:properRC}
\end{figure}

\index{filters} The rf electrodes have a small capacitive coupling $C_s$ to each control electrode
(typically less than $0.1$~pF), which can result in rf potential on the control electrodes.
This rf potential needs to be shunted to ground by a capacitor $C_f$ as shown in fig.~\ref{ions:fig:properRC}\index{rf shunt}.  
A low-pass RC filter
(typically $R=1~k\Omega$ and $C_f=1~nF$) on each control electrode is used to filter 
noise introduced by the externally-applied control electrode potentials.
The impedance of the lines between the control electrodes and $C_f$ should
be low or the rf shunting to ground will be compromised.
Proper grounding, shielding and filtering of the electronics
supplying the control electrode potentials are also important to suppress pickup and ground
loops (which can cause motional heating; see section~\ref{ions:sec:motional}).  

% %%%%%%%%%%%%%%%%%%%%%%%%%%%%%%%%%%%%%%%%%%%%%%%%%%%%%%%%%%%%%%%%%%%%%%%%%%%%%%
% %
% %%%%%%%%%%%%%%%%%%%%%%%%%%%%%%%%%%%%%%%%%%%%%%%%%%%%%%%%%%%%%%%%%%%%%%%%%%%%%%
\subsection{Motional heating}
\label{ions:sec:motional}\index{ions!motional heating}\index{heating}

Doppler and Raman cooling can place a trapped ion's harmonic motion into the ground state
with high probability \cite{ions:bible,ions:Diedrich89.PRL.62.403,ions:Monroe95.PRL.75.4011,ions:King98.PRL.81.1525}.
If we are to use the internal states of an ion to store information, we must turn off the
cooling laser beams during that period.  Unfortunately, the ions do not remain in the motional ground state and this 
heating can reduce the fidelity of operations performed with the ions. 
One source of heating comes from laser interactions used to manipulate the electronic
states \cite{ions:Ozeri07.PRA.75.042329}.  Another source is ambient electric fields that have a frequency 
component at the ion's motional frequencies. We expect such fields from the
\index{Johnson noise} Johnson noise on the electrodes \cite{ions:bible,ions:Turchette00.PRA.61.063418,ions:Deslauriers06.PRL.97.103007,ions:Leibrandt07.QIC.7.052}, 
but the heating rates observed experimentally
are typically several orders of magnitude larger than the Johnson noise can account for.
Currently, the source of this \index{anomalous heating} anomalous heating is not explained, but recent experiments 
\cite{ions:Deslauriers06.PRL.97.103007,ions:Labaziewicz08.PRL.100.013001} indicate it is thermally activated and consistent with patches of 
fluctuating potentials with a size scale smaller than the ion-electrode spacing \cite{ions:Turchette00.PRA.61.063418}.

\begin{figure}
   \centering
   \includegraphics{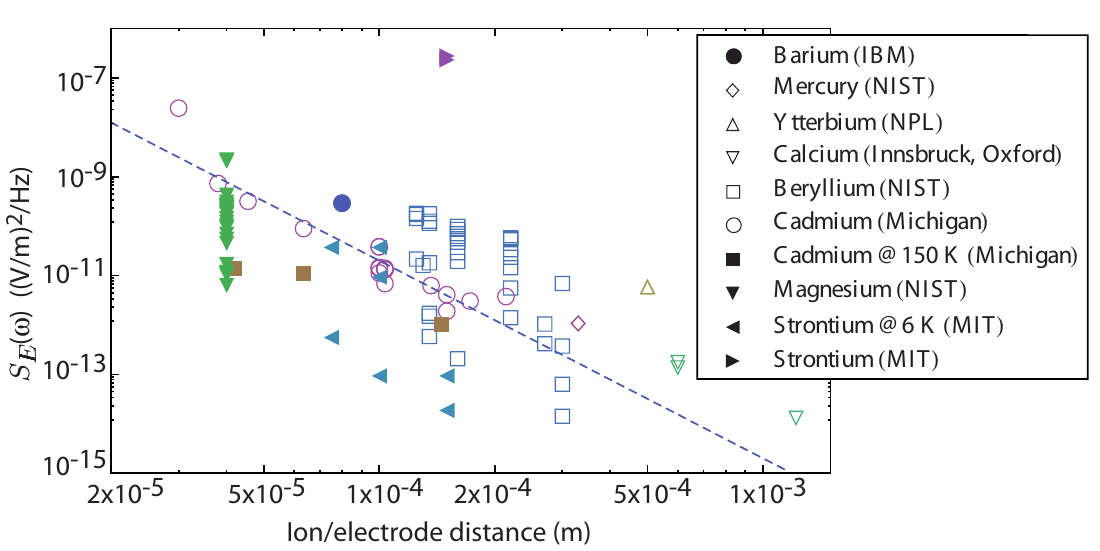}
   \Caption{Spectral density of electric-field fluctuations inferred from
  observed ion motional heating rates. Data points show heating measurements in
  ion traps observed in different ion species by several research groups  \cite{ions:Diedrich89.PRL.62.403,ions:Monroe95.PRL.75.4011,ions:Roos99.PRL.83.4713,ions:Tamm00.PRA.61.053405,ions:Turchette00.PRA.61.063418,ions:DeVoe02.PRA.65.063407,ions:Rowe02.QIC.2.257,ions:Deslauriers04.PRA.70.043408,ions:Deslauriers06.PRL.97.103007,ions:Home06.Thesis,ions:Stick06.NaturePhys.171.36,ions:Epstein07.PRA.76.033411,ions:Labaziewicz08.PRL.100.013001,ions:Britton08.Thesis,ions:blakestad2008a}.
  Unless specified, the data was taken with the trap at room temperature.
  The dashed line shows a $R^{-4}$ trend for ion heating vs ion-electrode
  separation $R$.}
   \label{ions:fig:ionHeatingScatterPlot}
 \end{figure}

The spectral density of electric field fluctuations $S_E$ at the ion's position
 inferred from ion heating measurements in a number of traps is plotted  
 versus the minimum ion-electrode separation $R$ in 
fig.~\ref{ions:fig:ionHeatingScatterPlot}. The dependence of $S_E$ on 
 $R$ and on the trap frequency $\omega$ follows a roughly $R^{-\alpha}\omega^{-\beta}$
scaling, where $\alpha\approx3.5$\cite{ions:Turchette00.PRA.61.063418,ions:Deslauriers06.PRL.97.103007} and $\beta\approx0.8$ to 1.4
\cite{ions:Turchette00.PRA.61.063418, ions:Deslauriers06.PRL.97.103007, ions:Seidelin06.PRL.96.253003,ions:Labaziewicz08.PRL.100.013001}.
In addition to being too small
to account for these measured heating rates,  Johnson noise 
scales as $R^{-2}$ \cite{ions:bible,ions:Turchette00.PRA.61.063418}. 
One candidate mechanism that does scale as $R^{-4}$
is noise caused by small fluctuating patch potentials on the electrode 
surfaces  \cite{ions:Turchette00.PRA.61.063418}. 
The potentials on these patches fluctuate at megahertz frequencies and generate a corresponding
fluctuating electric field at the ion's equilibrium position.
This field can lead to heating of the ion \cite{ions:Lamoreaux97.PRA.56.4970,ions:James98.PRL.81.317,ions:Henkel99.ApplPhysB.69.379,ions:Turchette00.PRA.61.063418, ions:Deslauriers06.PRL.97.103007,ions:Leibrandt07.QIC.7.052}. 

In the context of ion quantum information processing, microtraps are advantageous 
because quantum logic gate speeds and ion packing densities increase as the trap size decreases
\cite{ions:bible, ions:Kielpinksi02.Nature.417.709, ions:Leibfried03.Nature.422.412, ions:Kim05.QIC.5.515}. 
However, these gains are at odds with the
highly unfavorable dependence of motional heating on ion-electrode distance.  
For example, extrapolating from the room temperature heating results of \cite{ions:Seidelin06.PRL.96.253003},
 a $R=10$~\textmu m trap might exceed
$10^6$~quanta per second. Heating between gate operations 
can also be problematic because hot ions require more time to recool to the motional
ground state.

%%%%%%%%%%%%%%%%%%%%%%%%%%%%%%%%%%%%%%%%%%%%%%%%%%%%%%%%%%%%%%%%%%%%%%%%%%%%%%%
%%
%%%%%%%%%%%%%%%%%%%%%%%%%%%%%%%%%%%%%%%%%%%%%%%%%%%%%%%%%%%%%%%%%%%%%%%%%%%%%%%
\section{Measuring heating rates}
\label{ions:sec:recooling}\index{heating}

\begin{figure}
  \centering
  \includegraphics{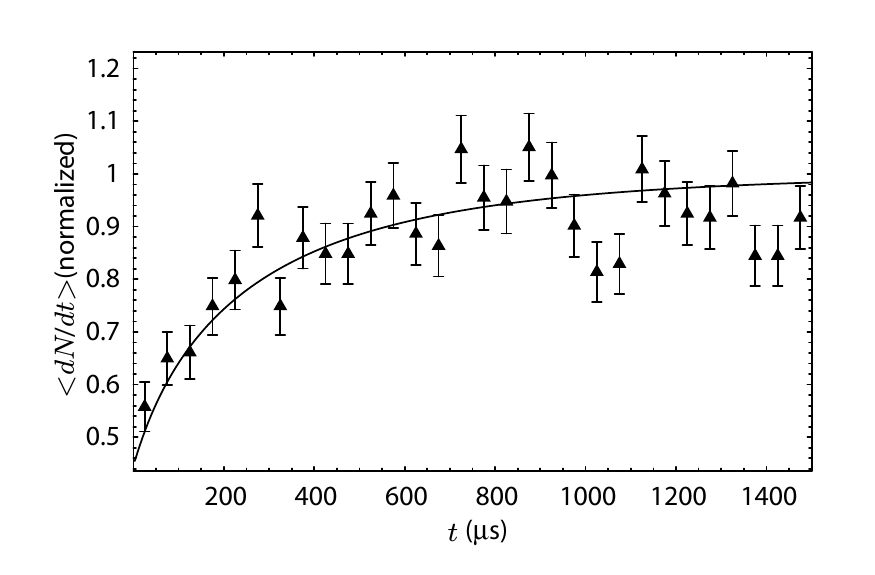}
  \Caption
  {Plot showing normalized fluorescence rate $dN/dt$ during Doppler cooling
  of a hot ion versus the time the cooling laser is turned off (dark time). 
  The experimental data averaged over many experiments is fit to the
  1D model \cite{ions:Wesenberg07.PRA.76.053416} briefly discussed in the text. The fit has a single free parameter:
  the ion's temperature at the outset of cooling. The error bars are
  based on counting statistics. Data taken on $^{25}Mg^+$ with a dark time of 25~s \cite{ions:Wesenberg07.PRA.76.053416}.}
  \label{ions:fig:recoolingSampleData}
\end{figure}

Heating rates have often been measured by observing an ion's energy increase 
after cooling to the motional ground state, a relatively complicated and 
technically challenging undertaking 
\cite{ions:Diedrich89.PRL.62.403,ions:Monroe95.PRL.75.4011}.
This section outlines a method to measure
ion motional heating with a single low power laser beam \cite{ions:Epstein07.PRA.76.033411,ions:Wesenberg07.PRA.76.053416,ions:Huber08.NJP.10.013004}.
Near resonance, an atom's fluorescence rate
is influenced by its motion due to the Doppler effect.
This can be exploited in the following way:
\begin{enumerate}
  \item Cool a trapped ion to its Doppler limit.
  \item Let it remain in the dark for some time. Ambient electric fields couple
  to the ion's motion and heat it.
  \item Turn on the Doppler cooling laser and measure the ion's time-resolved
  fluorescence, as shown in fig. \ref{ions:fig:recoolingSampleData}.
  \item A fit to a theoretical model of the ion fluorescence rate versus time (during recooling) \cite{ions:Wesenberg07.PRA.76.053416} 
  gives an estimate of the ion's temperature at the end of step 2.
\end{enumerate}

The theoretical model in \cite{ions:Wesenberg07.PRA.76.053416} explored cooling of hot ions
where the average modulus of  the Doppler shift is on the order of, or greater than, the cooling transition
line width $\Gamma$. The model is a one-dimensional semiclassical
theory of Doppler cooling in the weak binding limit where $\omega_{z}ll\Gamma$.
It is assumed that hot ions undergo harmonic oscillations with amplitudes
corresponding to the Maxwell-Boltzman energy distribution when averaged over many experiments.

As a one-dimensional (1D) model, only a single motional mode is assumed
to be hot. Since the electric field spectral density $S_{E}$ at the
ion is observed to scale approximately as $S_{E}\propto\omega^{-\beta}$,
where $\beta\approx0.8$ to 1.4 \cite{ions:Turchette00.PRA.61.063418, ions:Deslauriers06.PRL.97.103007, ions:Seidelin06.PRL.96.253003,ions:Labaziewicz08.PRL.100.013001}, the heating is effectively 1D if $\omega_{z}<<\omega_{x},\omega_{y}$. This is also important experimentally
because efficient Doppler cooling requires laser beam overlap with
all modes simultaneously: a change in ion fluorescence can arise from
heating of any mode. 

Heating rates measured with the \index{recooling} recooling technique were found to be in reasonable
agreement with rates measured starting from the ground state and allowing heating to 
only a few average motional quanta \cite{ions:Diedrich89.PRL.62.403,ions:Monroe95.PRL.75.4011}. In these 
comparisons, heating seems to be approximately linear from the ground state 
to at least  10000 motional quanta. The disadvantage of the recooling technique 
is that for small heating rates, the duration of step 2 can become quite long. 

%%%%%%%%%%%%%%%%%%%%%%%%%%%%%%%%%%%%%%%%%%%%%%%%%%%%%%%%%%%%%%%%%%%%%%%%%%%%%%%
%%
%%%%%%%%%%%%%%%%%%%%%%%%%%%%%%%%%%%%%%%%%%%%%%%%%%%%%%%%%%%%%%%%%%%%%%%%%%%%%%%
\section{Multiple trapping zones}
\label{ions:sec:multiplezones} \index{ion trap} \index{zones}

Much of the emphasis in the recent generation of ion traps is towards traps that can
store ions in multiple trapping zones and can transport ions \index{transport}
between the zones.

\begin{figure}[htb]
\centering
\includegraphics{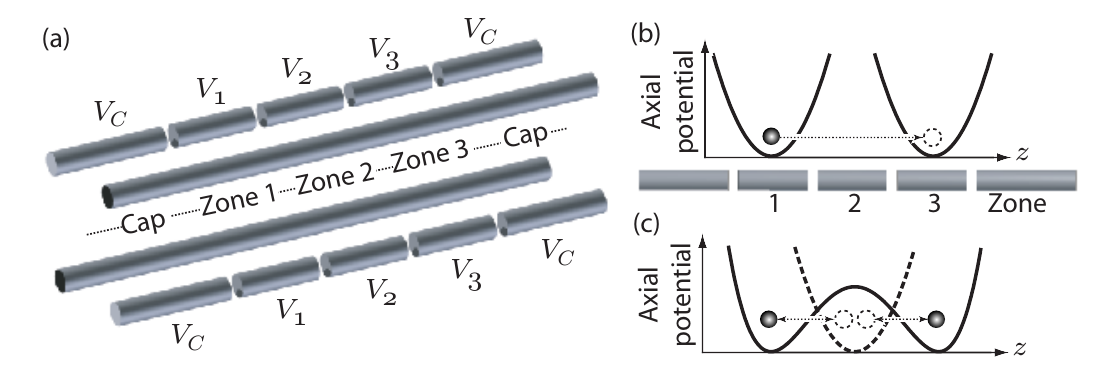}
\Caption{(a) Example of a multizone trap. By applying appropriate waveforms 
to the segmented control electrodes,
ions can be (b) shuttled from zone to zone or (c) pairs of ions
can be merged into a single zone or split into separate zones. } % Note the capital "C" in "\Caption"
\label{ions:fig:multizone}
\end{figure}

We can modify the basic Paul trap in fig.~\ref{ions:fig:LinEle} to support multiple zones and ion
transport by dividing the control electrodes into a series of segments as shown in
fig.~\ref{ions:fig:multizone}a. By applying appropriate potentials 
\cite{ions:Rowe02.QIC.2.257, ions:Barrett04.Nature.429.737, ions:Wineland05.ICOLS.393, ions:Home06.QIC.6.289,ions:Schulz07,ions:Reichle07,ions:Hucul08.QIC.8.501,ions:Huber08.NJP.10.013004} 
to these segments,
an axial harmonic well can be moved along the length of
the trap carrying ions along with it (fig.~\ref{ions:fig:multizone}b). 
In the adiabatic limit (with respect to $\omega_z^{-1}$), 
ions have been transported a distance of 1.2~mm in 50~\textmu s with 
undetectable heating  or internal-state decoherence \cite{ions:Rowe02.QIC.2.257}. 

As an example relevant to quantum information processing, we need to be able to take pairs of ions in a
single zone (for example, zone 2 in fig.~\ref{ions:fig:multizone}c)  and separate them into independent zones (one
ion in zone 1 and a second in zone 3) without excessive heating. Likewise, we need to reverse this
process and combine the ions into a single well.  Separating and recombining \index{separation}
are more difficult tasks than ion transport; the theory is discussed in \cite{ions:Home06.QIC.6.289} and
experimentally demonstrated in \cite{ions:Rowe02.QIC.2.257,ions:Barrett04.Nature.429.737}.
The basis for these potentials is the quadratic and quartic terms of
the axial potential. Proper design of the trap electrodes can increase the
strength of the quartic term and facilitate faster ion separation and merging
with less heating. Groups of two and three ions have been separated while heating 
the center of mass mode to less than 10 quanta and the higher order modes to less 
than 2 quanta \cite{ions:Barrett04.Nature.429.737}.   

The segmented Paul trap in fig.~\ref{ions:fig:multizone} forms a linear series
of trapping zones, but other
geometries are desirable.  Of particular interest are junctions with
linear trapping regions extending
from each leg.  Specific junction geometries are discussed in section~\ref{ions:sec:traps}. 
The broad goal is to create large interconnected trapping structures
that can store, transport and reorder ions so that any two ions can be brought
together in a common zone \cite{ions:bible,ions:Kielpinksi02.Nature.417.709}.

%%%%%%%%%%%%%%%%%%%%%%%%%%%%%%%%%%%%%%%%%%%%%%%%%%%%%%%%%%%%%%%%%%%%%%%%%%%%%%%
%%
%%%%%%%%%%%%%%%%%%%%%%%%%%%%%%%%%%%%%%%%%%%%%%%%%%%%%%%%%%%%%%%%%%%%%%%%%%%%%%%
\section{Trap modeling}
\label{ions:sec:modeling}\index{modeling}

Calculation of trap depth, secular frequencies, and transport and separation
waveforms requires detailed knowledge of the potential and electric fields near
the trap axis. In the pseudopotential approximation, the general time-dependent
problem is simplified to a slowly varying electrostatic one.  For simple four-rod type traps, good
trap design is not difficult using numerical simulation owing to their symmetry.
However, SE trap design is more complicated since the potential may have large
anharmonic terms and highly asymmetric designs are common.
Fortunately, for certain SE trap geometries, analytic solutions exist. These closed-form expressions
permit efficient parametric optimization of electrode geometries not practical by
numerical methods.  In this section, we will first discuss the full 3D
calculations and then introduce the analytic solutions.

%%%%%%%%%%%%%%%%%%%%%%%%%%%%%%%%%%%%%%%%%%%%%%%%%%%%%%%%%%%%%%%%%%%%%%%%%%%%%%%
%% 
%%%%%%%%%%%%%%%%%%%%%%%%%%%%%%%%%%%%%%%%%%%%%%%%%%%%%%%%%%%%%%%%%%%%%%%%%%%%%%%
\subsection{Modeling 3D geometries}
\label{ions:sec:modelinggeom}

There are several numerical methods for solving the general electrostatic
problem. In our trap simulations, we use the boundary element method
implemented in a commercial software package.  In contrast to the finite
element method,  the solutions from the boundary element method are 
in principle differentiable to all orders.  A simulation consists of calculating
the potential due to each control electrode when that electrode is set to a fixed non-zero potential
and all others are grounded. The
solution for an arbitrary set of potentials on the control electrodes 
is then a linear combination of these particular solutions.
Similarly, the pseudopotential is obtained by scaling the field calculated
for a finite potential on the rf electrodes and ground on the control electrodes
 and then squaring the field according to eq. (\ref{ions:pseudopotential}).

%%%%%%%%%%%%%%%%%%%%%%%%%%%%%%%%%%%%%%%%%%%%%%%%%%%%%%%%%%%%%%%%%%%%%%%%%%%%%%%
%%
%%%%%%%%%%%%%%%%%%%%%%%%%%%%%%%%%%%%%%%%%%%%%%%%%%%%%%%%%%%%%%%%%%%%%%%%%%%%%%%
\subsection{Analytic solutions for surface electrode traps}
\label{ions:sec:analyticsolutions}\index{surface electrode (SE) trap}\index{analytic solutions}

\begin{figure}[htb]
\centering
\includegraphics{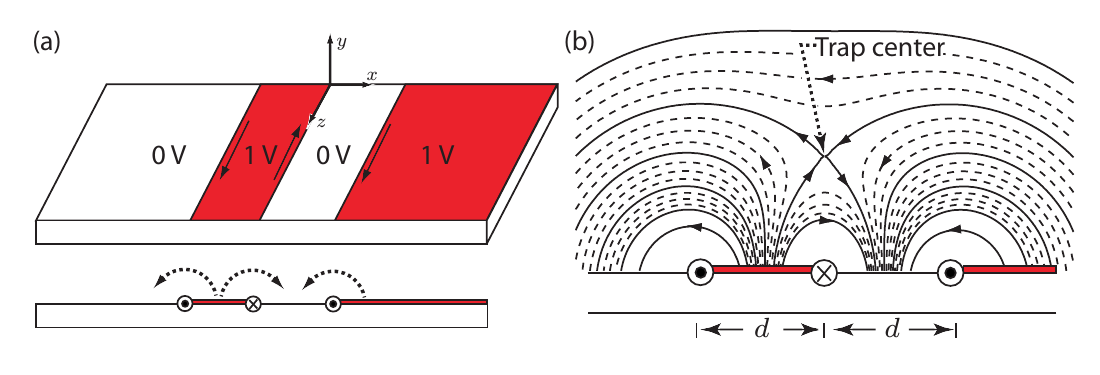}
\Caption{Surface electrode trap composed of two rf electrodes embedded in
a ground plane (four-wire trap) (a).  The field lines from the Biot-Savart type integral 
are shown in (b).} 
% Note the capital "C" in "\Caption"
\label{ions:fig:lincalc}
\end{figure}

Numerical calculations work for any electrode geometry, but they are are slow and
not well suited to automatic optimization of SE trap electrode shapes. For the
special case of SE traps, an analytic solution exists subject to a few realistic
geometric constraints. Electrodes are modeled as a collection of separately biased regions
embedded in an
infinite ground plane (see fig.~\ref{ions:fig:lincalc}) without gaps between
the electrodes.  The
electric field that would be observed from a biased region is proportional to the
magnetic field produced by a current flowing along its perimeter
~\cite{ions:Oliveira01.EJP.22.31}. The problem is then reduced from
solving Laplace's equation to integrating a Biot-Savart type integral 
around the patch boundary. Furthermore, for patches that have boundaries composed of straight
line segments, the integrals have analytic solutions. The application of this technique
to SE traps is given in \cite{ions:wesenberg2008b}.

The main shortcoming of this method is the requirement that there be no
gaps between the electrodes.  Typical SE trap fabrication techniques produce
1 to 5~{\textmu m} gaps which can only be accounted for at the level important
to ion dynamics by full numerical simulations.

Fields for arbitrarily shaped patches can be calculated using
this Biot-Savart technique, but for simplicity we restrict ourselves to strips that extend to
infinity in the $z$-direction of fig.~\ref{ions:fig:lincalc}. For this particular case, we can also derive potentials from the
calculated fields. A strip extending from $x=a$ to $x=b$ with $a<b$ held at potential
$U_s$ leads to a spatial potential
\begin{equation}
\label{ions:InfStr}
	\Phi_s(a,b)=\frac{U_s}{\pi}\times\left\{
		\begin{array}{ll}
			\tan^{-1}\left(\frac{x-a}{y}\right)-\tan^{-1}\left(\frac{x-b}{y}\right),
					&-\infty<a<b<\infty \\
			\frac{\pi}{2}-\tan^{-1}\left(\frac{x-b}{y}\right),
					& a=-\infty \\
			\frac{\pi}{2}+\tan^{-1}\left(\frac{x-a}{y}\right).
				& b=\infty\\
		\end{array}
\right.
\end{equation}
The potentials of multiple, non-overlapping strips can then be summed for more complex structures.

\begin{figure}[htb]
\centering
\includegraphics{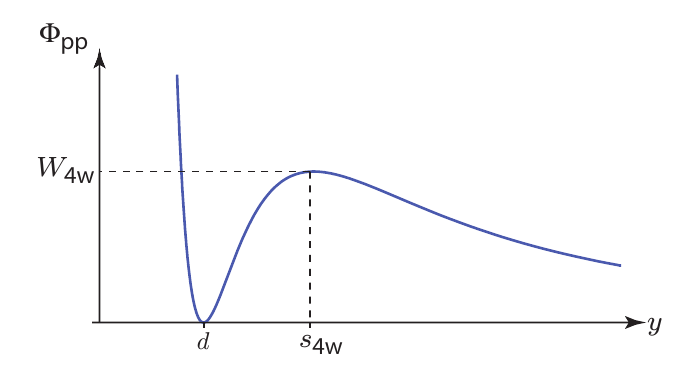}
\Caption{Analytic pseudopotential of the four-wire trap along $y$ at $x=0$. The
trapping zero is at $y=d$; the maximum defining the well depth is at $s_{\rm 4w}=
d~\sqrt{2+\sqrt{5}}$} % Note the capital "C" in "\Caption"
\label{ions:fig:pseupot4wire}
\end{figure}

Two basic SE trap geometries are the \index{four-wire trap} `four-wire' trap and the \index{five-wire trap} `five-wire' trap. An example
four-wire trap consists of an rf electrode from $x=-d$ to $x=0$ and another
semi-infinite rf electrode from $x=d$ to $x=\infty$ (see
fig.~\ref{ions:fig:lincalc}a and fig.~\ref{ions:fig:planar}). An example five-wire trap consists of two symmetric
rf electrodes from $x= -3/2 d$ to $x=-1/2 d$ and $x=1/2 d$ to $x=3/2 d$. Their
respective potentials are given by
\begin{equation}
\label{ions:FouWir}
  \Phi_{\rm 4w}=\Phi_s(-d,0)+\Phi_s(d,\infty);
  ~~\Phi_{\rm 5w}=\Phi_s\left(-\frac{3 d}{2} ,
  -\frac{d}{2} \right)+\Phi_s\left(\frac{d}{2},
   \frac{3 d}{2}\right).
\end{equation}
From the electric fields and eq. (\ref{ions:pseudopotential}) we can derive the
pseudopotential. Note that the potential minima coincide with the points of zero electric
field that lie in the line of symmetry around $x=0$ at $y_{4w}=d$ and
$y_{5w}=\sqrt{3}d/2$, respectively. For an ion of mass $m$ and charge $q$, 
the trap frequencies along the two degenerate radial directions are
\begin{equation}
\label{ions:TraFre}
\omega_{\rm 4w}= \frac{q U_s}{\sqrt{2} m \pi \Omega_{\rm rf} d^2 };
~~\omega_{\rm 5w}=\sqrt{\frac{2}{3}}\frac{q U_s}{m \pi \Omega_{\rm rf} d^2 },
\end{equation}
where $\Omega_{\rm rf}$ is the rf-drive frequency. 
Figure~\ref{ions:fig:pseupot4wire} shows the general shape of the
pseudopotential well along the $y$-axis at $x=0$ for the four-wire trap (for
the five-wire trap the potential looks very similar).
The potential is zero at $y=d$ where the ion is trapped,
then rises to a maximum and finally asymptotically drops towards zero for
$y \rightarrow \infty$. The positions of the maxima are at
\begin{equation}
\label{ions:SadPoiPos}
s_{\rm 4w}= d~\sqrt{2+\sqrt{5}};~~s_{\rm 5w}=d~\sqrt{3/4+\sqrt{3}},
\end{equation}
and the pseudopotential well depth (in eV) is
\begin{equation}
\label{ions:SadPoiDepth}
W_{\rm 4w}= \left(\frac{q U_s^2}{4 m \Omega_{\rm rf}^2}\right) \frac{2}{\pi^2 d^2(11+5\sqrt{5})};
~~W_{\rm 5w}=\left(\frac{q U_s^2}{4 m \Omega_{\rm rf}^2}\right) \frac{1}{\pi^2 d^2(7+4\sqrt{3})}.
\end{equation}
To get an idea of practical parameters, we can calculate the radial
frequency and pseudopotential well depth of a four-wire trap with a geometry
similar to the trap described in \cite{ions:Seidelin06.PRL.96.253003}. For 
$\Omega_{\rm rf}/2 \pi=  87$ MHz, $U_s=103.2$ V, $d=$ 40~\textmu m and $m$ the mass of a $^{24}$Mg$^+$
ion, we get $\omega_{\rm 4w}/2 \pi =$16.9 MHz and $W_{\rm 4 w}=$203 meV. 

%%%%%%%%%%%%%%%%%%%%%%%%%%%%%%%%%%%%%%%%%%%%%%%%%%%%%%%%%%%%%%%%%%%%%%%%%%%%%%%
%%
%%%%%%%%%%%%%%%%%%%%%%%%%%%%%%%%%%%%%%%%%%%%%%%%%%%%%%%%%%%%%%%%%%%%%%%%%%%%%%%
\section{Trap examples}
\label{ions:sec:traps}

Having covered the general principles for Paul trap designs,  we now give
specific examples of microfabricated ion traps.  A number of fabrication
techniques have been used for micro-traps, starting with assembling multiple
wafers to form a traditional Paul trap type design
\cite{ions:Rowe02.QIC.2.257,ions:Wineland05.ICOLS.393,ions:Hensinger06.APL.88.034101,ions:Huber08.NJP.10.013004,ions:Schulz08.NJP.10.045007,ions:Britton08.Thesis,ions:blakestad2008a}.
Recently, trap fabrication
has been extended to monolithic designs using substrate materials such as Si,
GaAs, quartz, and printed circuit board \cite{ions:Pearson06.PRA.73.032307,ions:Seidelin06.PRL.96.253003,ions:Britton06.ArXiv.0605170, ions:Stick06.NaturePhys.171.36,ions:Pau06.PRL.96.120801,ions:Brown07.PRA.75.015401, ions:Labaziewicz08.PRL.100.013001,ions:Britton08.Thesis}.  
The fabrication process includes such microfabrication
standards as photolithography, metalization, and chemical vapor deposition as
well as other less used techniques such as laser machining.

\begin{figure}[htb]
\centering
\includegraphics{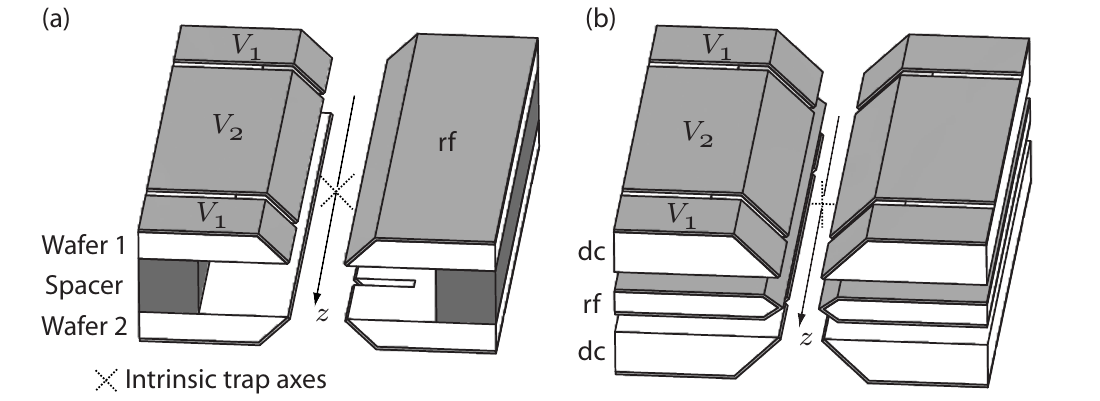}
\Caption{Multiwafer traps can be formed by mechanically clamping or bonding multiple substrates to
form (a) a four-rod quadrupolar Paul trap type structure or (b) a modified Paul trap using a three-layer structure
\cite{ions:Hensinger06.APL.88.034101}. The segmentation of the control electrodes on
the bottom substrate is similar to that of the top substrate.} % Note the capital "C" in "\Caption"
\label{ions:fig:multiwafer}
\end{figure}

\begin{figure}[htb]
\centering
\includegraphics{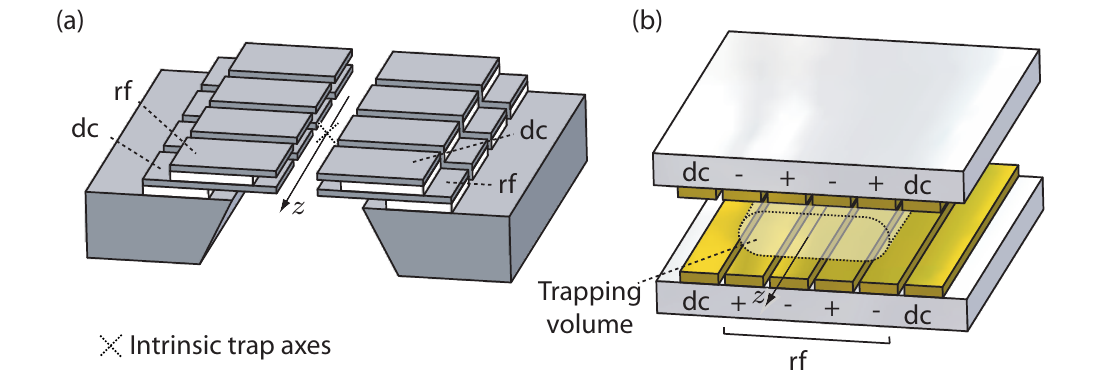}
\Caption{(a) Four-rod Paul trap realized by successively deposited layers of GaAs and AlGaAs 
on a GaAs wafer \cite{ions:Stick06.NaturePhys.171.36}.
In (b), conducting gold strips deposited on two glass substrates and alternately driven 
at opposite phases of an rf  source (phases denoted by `+' and `-') generates a trapping 
volume between the substrates \cite{ions:Debatin08.PRA.77.033422}. 
Static potentials at the edges of the trap along the $z$ axis, applied with electrodes
that are not shown, confine the ions to the central region of the trap.} 
% Note the capital "C" in "\Caption"
\label{ions:fig:misctraps}
\end{figure}

\index{junction}The microfabricated equivalent to the prototypical four-rod Paul trap can use
two insulating substrates patterned with electrodes that are then 
clamped or bonded together with an insulating spacer.  
This approach has been implemented in a number of traps 
~\cite{ions:Rowe02.QIC.2.257,ions:Barrett04.Nature.429.737,ions:Wineland05.ICOLS.393, ions:Britton08.Thesis,ions:Schulz08.NJP.10.045007,ions:blakestad2008a}
using two substrates, as shown in fig.~\ref{ions:fig:multiwafer}a.
Alternatively, it is possible to build this structure into a single 
monolithic device \cite{ions:Stick06.NaturePhys.171.36}, as indicated 
schematically in fig.~\ref{ions:fig:misctraps}a.

\begin{figure}[htb]
\centering
\includegraphics{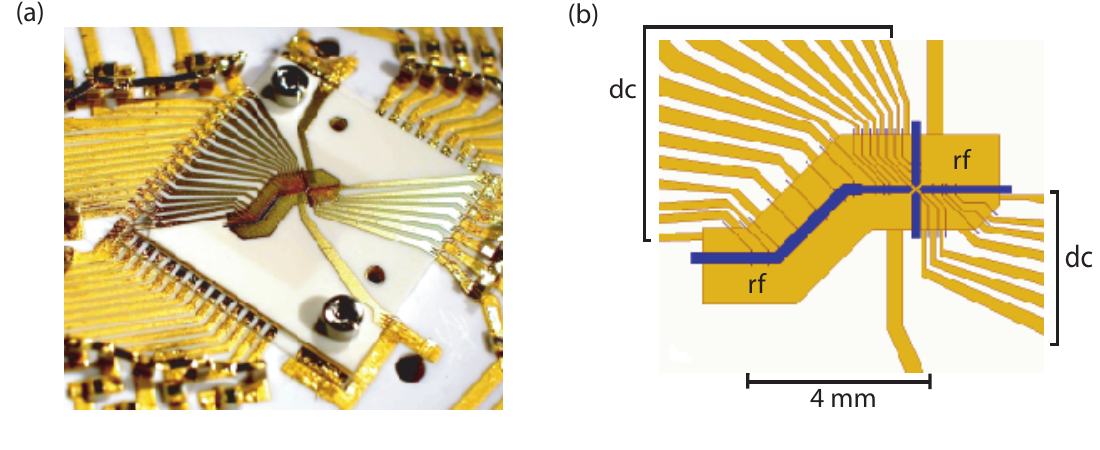}
\Caption{Example of a two-wafer trap with an `X' junction \cite{ions:blakestad2008a}. The trap 
electrodes are fabricated with 
evaporated and electroplated gold 
that is deposited on laser-machined alumina substrates.} 
% Note the capital "C" in "\Caption"
\label{ions:fig:brad}
\end{figure}

Reference \cite{ions:Hensinger06.APL.88.034101} describes
a three-wafer trap design like that shown in fig.~\ref{ions:fig:multiwafer}b
incorporating a `T' shaped junction. At NIST, a two-layer trap with an `X' junction 
has recently been implemented \cite{ions:blakestad2008a} and is shown in
fig.~\ref{ions:fig:brad}.  Such two-dimensional 
geometries will be important in order to combine arbitrarily 
selected qubits from an array together in the same trap zone.

Another approach demonstrated recently  used
two patterned substrates, without slots, that are
 mounted with the conducting layers facing each other
 \cite{ions:Debatin08.PRA.77.033422} (see
fig.~\ref{ions:fig:misctraps}b). The array of conducting gold electrode strips
 is driven with rf that alternates between a phase of $0\,^\circ$ and $180\,^\circ$ from one strip 
 to the next. This creates a pseudopotential that is near
zero for much of the space between the wafers but which rises sharply near the
substrates. When combined with static potentials at the edges of the wafers,
this trap generates a near field-free region bounded by `hard' potential walls (fig.~\ref{ions:fig:misctraps}b).
Arrays of cylindrical Paul type traps have been microfabricated on silicon for use as
mass spectrometers \cite{ions:Pau06.PRL.96.120801}.

\begin{figure}[htb]
\centering
\includegraphics{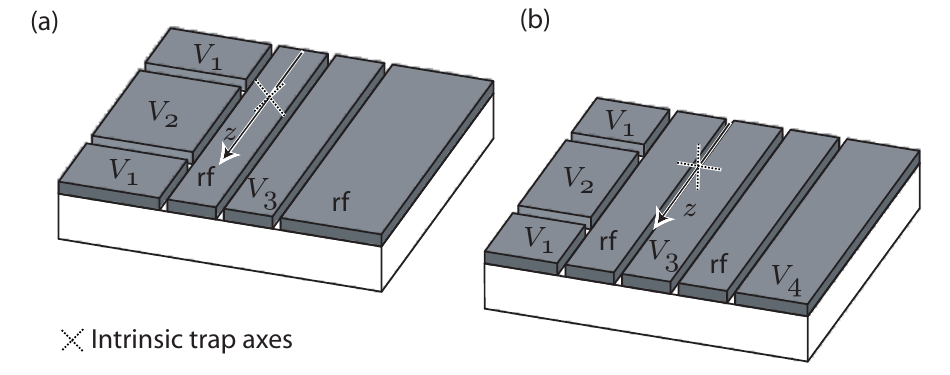}
\Caption{(a) Four-wire SE trap geometry and (b) symmetric five-wire SE trap
geometry.  In practice, the symmetric five-wire geometry is typically not used because
of the difficulty of cooling the vertical motion of the trapped ions.} 
% Note the capital "C" in "\Caption"
\label{ions:fig:planar}
\end{figure}

\index{surface electrode (SE) trap}Surface electrode (SE) traps have the benefit of using standard microfabrication
methods where layers of metal and insulator are deposited on the surface of the
wafer without the need for milling of the substrate itself. There are two
general versions of the surface trap electrode geometry, as described in section 
\ref{ions:sec:analyticsolutions} and shown in
fig.~\ref{ions:fig:planar}.   The four-wire geometry has the intrinsic trap
axes rotated at 45$^\circ$ to the substrate plane, which allows for efficient laser
cooling of the ion. The five-wire geometry has one intrinsic trap axis perpendicular to
the surface, which can make that axis difficult to Doppler cool (see section~\ref{ions:sec:doppler}). 
To enable Doppler cooling, additional control electrodes can be added to the design to rotate the 
trap axes away from the intrinsic direction. Alternately, a hybrid between
the four- and five-wire designs where the rf strips are of unequal widths (an `asymmetric' five-wire trap)
will rotate the intrinsic axes and enable cooling.

Surface electrode traps are relatively new and only a few designs have been
demonstrated 
\cite{ions:Pearson06.PRA.73.032307,ions:Seidelin06.PRL.96.253003,ions:Britton06.ArXiv.0605170,ions:Brown07.PRA.75.015401, ions:Labaziewicz08.PRL.100.013001}. 
An SE trap was first demonstrated with charged polystyrene balls using 
standard PC board fabrication techniques \cite{ions:Pearson06.PRA.73.032307}.
The first SE trap for atomic ions was constructed on a fused quartz substrate
with electroplated gold electrodes \cite{ions:Seidelin06.PRL.96.253003,ions:Britton08.Thesis}.  
In addition, meander-line resistors were
fabricated on the chip as part of the control electrode filtering. Surface-mount capacitors
were gap welded to the chip to complete the filters (see section~\ref{ions:sec:elecInterconnect}).  The
fabrication process sequence is shown in fig.~\ref{ions:fig:signefab}. The bonding pads and
the thin meander-line resistors were formed by liftoff of evaporated gold.
Charging of the exposed substrate between the electrodes was a concern, so the
trap electrodes were made of 6~\textmu m thick electroplated gold with 8~\textmu m
gaps so as to shield the ion somewhat from the charges on the quartz surface.

\begin{figure}[htb]
\centering
\includegraphics{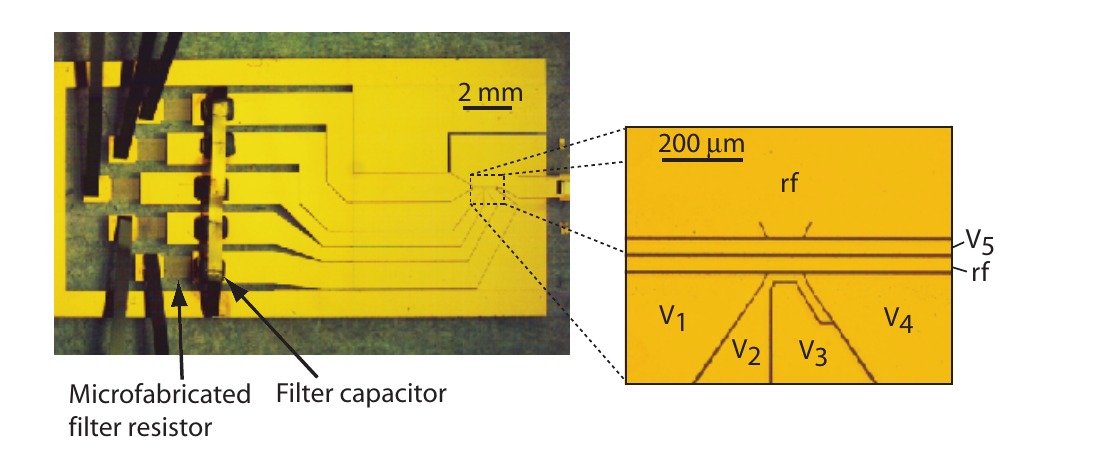}
\Caption{An example of a four-wire SE trap constructed of
electroplated gold on a quartz substrate \cite{ions:Seidelin06.PRL.96.253003}.} % Note the capital "C" in "\Caption"
\label{ions:fig:signe}
\end{figure}

\begin{figure}[htb]
\centering
\includegraphics{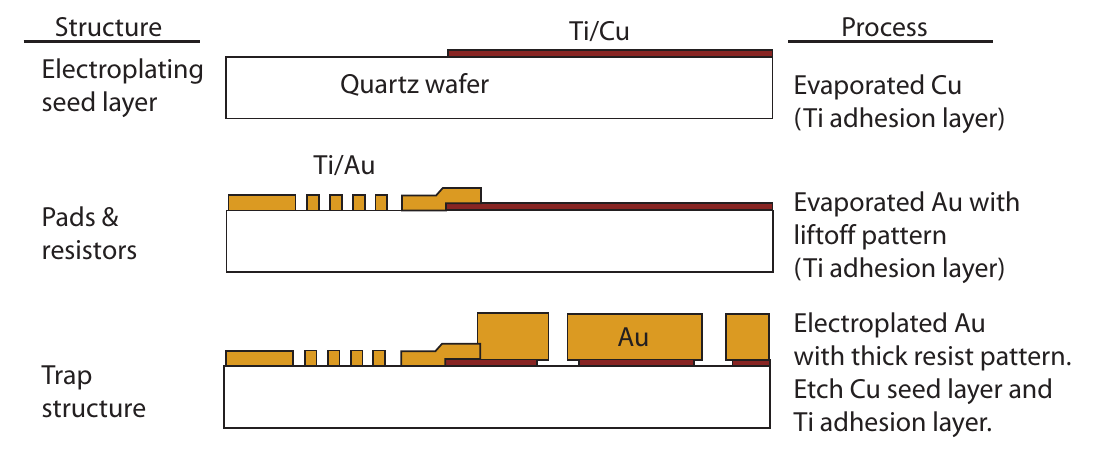}
\Caption{Fabrication steps for the example SE trap in fig.~\ref{ions:fig:signe} 
\cite{ions:Seidelin06.PRL.96.253003}.
The copper seed layer could not be used under the meander line resistors because the
final step of etching the seed layer would fully undercut the narrow meander
pattern.} % Note the capital "C" in "\Caption"
\label{ions:fig:signefab}
\end{figure}

A similar design  was built by a group at MIT for low-temperature testing 
using 1~\textmu m evaporated silver on quartz \cite{ions:Labaziewicz08.PRL.100.013001}. They
reported a strong dependence of the anomalous ion heating on temperature (see section~\ref{ions:sec:motional}).

The construction of the traps in \cite{ions:Seidelin06.PRL.96.253003} 
and \cite{ions:Labaziewicz08.PRL.100.013001} was based on adding conducting layers to an
insulating substrate.  An alternate fabrication method used boron-doped Si wafers
anodically bonded to a glass substrate \cite{ions:Britton06.ArXiv.0605170} and 
boron-doped silicon-on-insulator (SOI) wafers \cite{ions:Britton08.Thesis}. In both cases trenches were etched
through the silicon layer to the glass or embedded insulating layer to define the trap electrodes. The SOI
design demonstrated multiple trapping zones in a SE trap and backside loading of ions.

\begin{figure}[htb]
\centering
\includegraphics{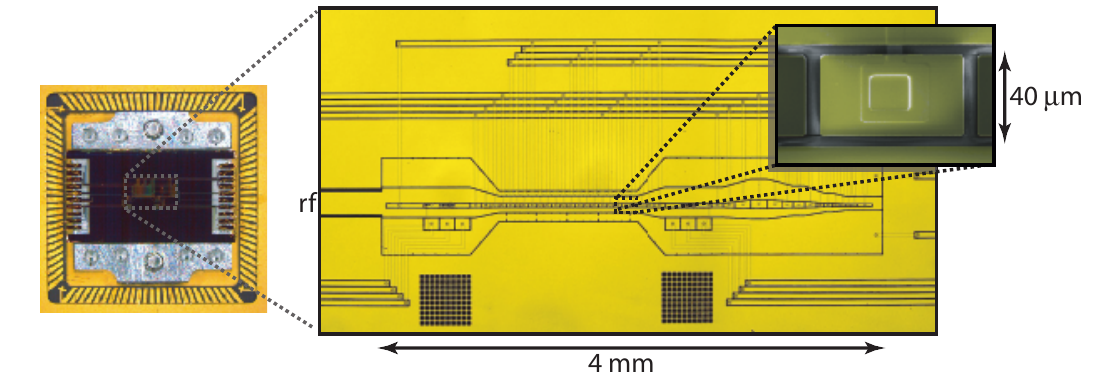}
\Caption{Multilayer, multi-zone, linear SE trap mounted in its carrier and an enlargement of the
active region \cite{ions:Amini07.BAPS}.} % Note the capital "C" in "\Caption"
\label{ions:fig:multilayer}
\end{figure}

\begin{figure}[htb]
\centering
\includegraphics{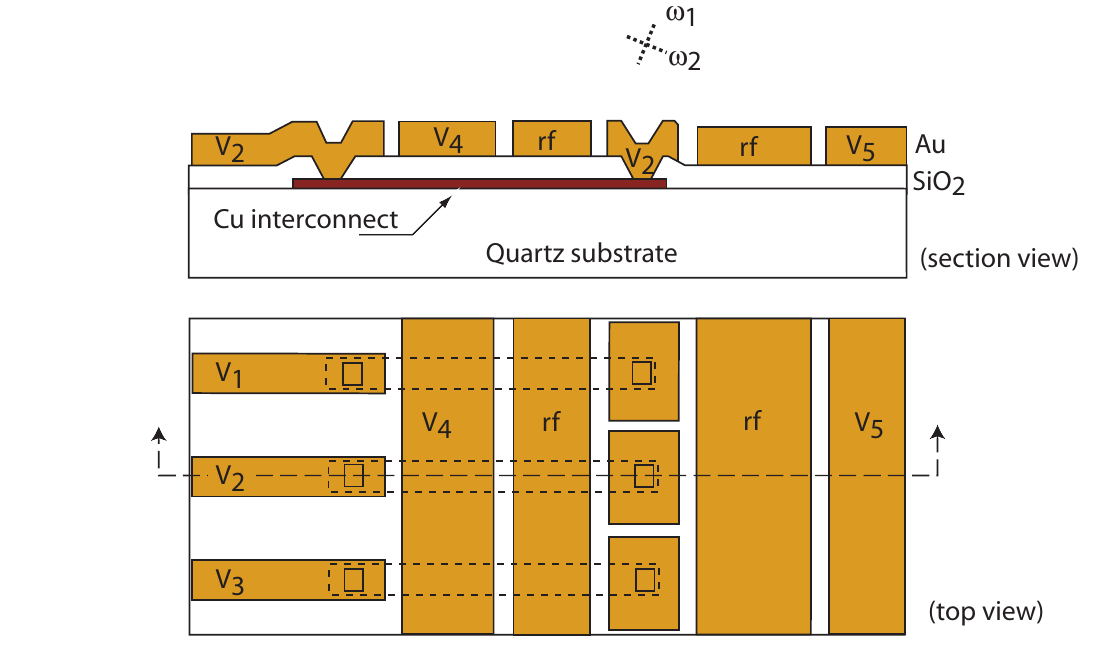}
\Caption{Fabrication of an asymmetric five-wire multilayer SE trap.  A CVD 
oxide insulates the surface electrodes from the second layer of
interconnects.  Plasma etched holes in the insulated layer connect the two
conducting layers.} 
% Note the capital "C" in "\Caption"
\label{ions:fig:multilayerfab}
\end{figure}

Surface electrode traps allow for complex arrangements of trapping zones, but
making electrical connections to these electrodes quickly becomes intractable
as the complexity grows.  This problem can be addressed by incorporating multiple
conducting layers into the design with only the field from the top layer
affecting the ion \cite{ions:Kim05.QIC.5.515,ions:Amini07.BAPS}. An example of such a
multilayer trap fabricated on an amorphous quartz substrate at NIST is shown in 
fig.~\ref{ions:fig:multilayer}. The metal layers are separated by
chemical vapor deposited (CVD) silicon dioxide and connections between metal layers are
made by vias that are plasma etched through the oxide, as shown in
fig.~\ref{ions:fig:multilayerfab}.  The fabrication process for the surface
gold layer is similar to the electroplating shown in
fig.~\ref{ions:fig:signefab}.

In the last three years, microfabricated traps have also been produced by 
Sandia National Laboratory (contact: M. Blaine, SNL) and Lucent Technologies 
(contact: R. Slusher, Georgia Tech Research Institute) and distributed to several ion trap groups in the 
framework of a "trap foundry" initiated by DTO (now IARPA). Several 
groups have seen trapping in the Lucent trap, a 17-zone SE trap. The 
Sandia trap, a 5-zone planar trap where the ions reside in-plane with 
the electrodes, has also been used to trap ions in two laboratories. 

%%%%%%%%%%%%%%%%%%%%%%%%%%%%%%%%%%%%%%%%%%%%%%%%%%%%%%%%%%%%%%%%%%%%%%%%%%%%%%%
%%
%%%%%%%%%%%%%%%%%%%%%%%%%%%%%%%%%%%%%%%%%%%%%%%%%%%%%%%%%%%%%%%%%%%%%%%%%%%%%%%
\section{Future}
\label{ions:sec:future}

As ion traps become smaller, trap complexity increases and
features such as junctions promise to expand the capabilities of
such traps. The two experimentally demonstrated atomic ion traps 
with junctions (see \cite{ions:Hensinger06.APL.88.034101} and fig.~\ref{ions:fig:brad}) 
are based on multilayer designs. The slots and difficulty of alignment and bonding in multiwafer 
traps make it difficult to scale such structures.  

\begin{figure}[htb]
\centering
\includegraphics{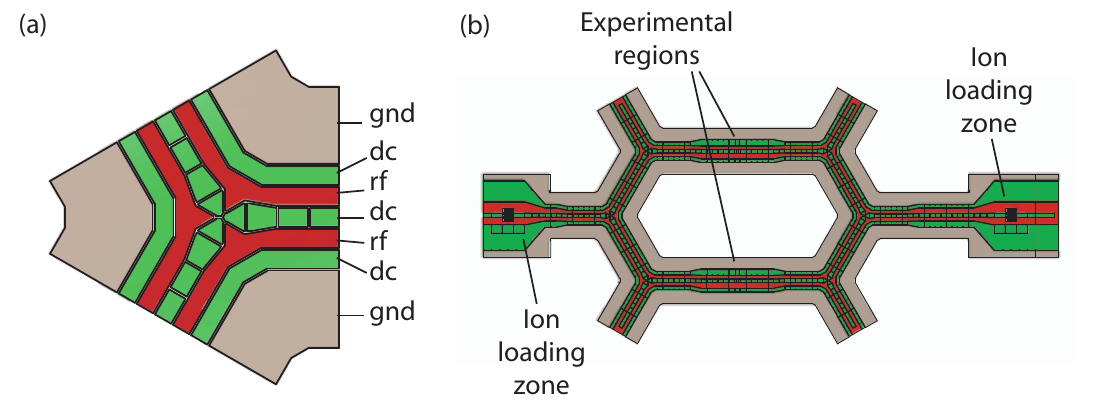}
\Caption{(a) Example of a SE trap `Y' junction and (b) a prototype design using
multiple `Y' junctions to link experimental regions and loading zones.} 
% Note the capital "C" in "\Caption"
\label{ions:fig:setjunction}
\end{figure}

\index{junction}Figure~\ref{ions:fig:setjunction}a shows an example design of a `Y' version of an 
SE trap junction. The shape of the rf 
junction is an example of the optimization that is possible with
SE traps because of the efficient methods described in section~\ref{ions:sec:analyticsolutions} 
to calculate the fields. The electrode geometry has been optomized
to generate a pseudopotential that has minimal 
axial `bumps' so that rf micromotion during ion transport will be minimized
(see section~\ref{ions:sec:micromotion}).  Components such as this `Y' could then be assembled
into larger structures as shown in fig.~\ref{ions:fig:setjunction}b.
Surface electrode traps fabricated using standard recipes in a foundry and using standard patterns
may eventually make ion traps more accessible to research groups that do not have the 
 resources needed to develop their own.

With increased trap complexity, several other issues arise. One of these 
is the question of how to package traps and provide all the electrical
connections needed to operate them. Another issue is that of corresponding 
complexity of the lasers used in 
manipulating the ions. Beyond cooling, state preparation, and detection, lasers are needed to
 coherently manipulate the internal states of the ions and couple
pairs or groups of ions.  Multiplexing sets of
lasers to address multiple trapping zones for 
parallel processing will be difficult. Alternatives
to laser optical field  state manipulation  have been  proposed 
\cite{ions:Mintert01.PRL.87.257904,ions:Leibfried07.PRA.76.032324,ions:Johanning07.arxiv.0801.0078, ions:Chiaverini08.PRA.77.022324, ions:Ospelkaus08.PRL.101.090502} where
magnetic structures, both active wire loops and passive magnetic layers,
replace laser beams. If proven to be viable, this would transfer much of the
experimental complexity from large laser systems to electronic packages, which
can be more reliably engineered and should be scalable \cite{ions:Kim05.QIC.5.515}.

\section{Acknowledgments}
\label{ions:sec:ack}

Work supported by the NIST Quantum Information Program and IARPA. 
This manuscript is a publication of NIST and is not subject to U.S. copyright.

\clearpage

\setlength{\bibindent}{4mm} % indentation for two digits

%===============================================================================
% Bibliography
%===============================================================================
\addcontentsline{toc}{section}{References}
\bibliographystyle{atchip}
\bibliography{ions} % your reference database

%===============================================================================
% Index
%===============================================================================
\newpage\addcontentsline{toc}{section}{Index}
\printindex

\end{document}